\documentstyle[aps,pre,psfig,floats]{revtex}
\begin{document}
\draft

\twocolumn[\hsize\textwidth\columnwidth\hsize\csname @twocolumnfalse\endcsname

\title{Novel surface state in a class of incommensurate systems\\}

\author{A. E. Jacobs$^{(1,2)}$, D. Mukamel$^{(2)}$ and D. W. Allender$^{(3)}$}

\address{$^{1}$ Department of Physics, University of Toronto, Toronto,
Ontario, Canada M5S 1A7 \\ $^{2}$
         Department of Physics of Complex Systems, The Weizmann Institute 
         of Science, Rehovot 76100, Israel\\ $^{3}$
	 Department of Physics and Liquid Crystal Institute, Kent State
	 University, Kent, Ohio 44242 USA \\ 
	 [-2mm] $ $}

\date{\today}

\maketitle
\begin{abstract}

We study the Landau model of the class of incommensurate systems with a 
scalar order parameter where the modulated phase is driven by a 
gradient-squared term with negative coefficient. 
For example, theoretical studies of cholesteric liquid crystals in a field 
(electric or magnetic) suggest that such an modulated phase should exist 
at high chirality.
The bulk phase diagram in the presence of a bulk external field which 
couples linearly to the order parameter exhibits a modulated phase inside 
a loop in the temperature-field plane, and a homogeneous phase outside.
On analyzing the same model for a semi-infinite system, we find a surprising 
result; 
the system exhibits surface states in a region where the bulk phase is 
homogeneous (but close to the modulated region). 
These states are very different from the well-known surface states induced 
either by a surface field or by enhanced interactions at the surface, for 
they exist and are energetically favored even when the sole effect of the 
surface is to terminate the bulk, as expressed by free boundary 
conditions taken at the surface.
Near the surface, the surface-state order parameter is very different from 
the bulk value (in fact, it has the opposite sign). 
When the temperature or the bulk field are varied to move away from the 
modulated state, we find a surface phase transition at which the surface 
states become energetically unfavorable, though they continue to exist as 
metastable states. 
We then study how a surface field changes the surface phase diagram. 
\end{abstract}
\vspace{2mm}
\pacs{PACS numbers: 61.30.Cz; 64.60.Kw; 64.70.Md}]


\section{Introduction}

It is well known that a surface field can give rise to wetting phenomena 
and also that enhanced interactions near a surface can give rise to surface 
order without bulk order \cite{Dietrich}. 
Nakanishi and Fisher \cite{Nakanishi} have given a unified picture of wetting 
and surface ordering at the phenomenological (Landau-theory) level; 
these effects require that surface terms be added to the bulk free energy. 
In this article, we report an entirely new surface effect which should occur
in a particular class of incommensurate systems. 
We find that surface states exist and are energetically favored by the mere 
presence of the surface, without surface terms like those considered in 
\cite{Nakanishi}. 
 
Candidate physical systems for observing these states include highly chiral 
cholesterics in electric or magnetic field, where a bulk undulating phase 
was recently predicted to occur \cite{SMA}. 
This phase is an undulating structure in which the amount of orientational 
order varies periodically in conjunction with an oscillation of the 
direction of the local optic axis.  
It is expected to occur under appropriate conditions of temperature and a 
strong aligning electric or magnetic field. 
As discussed in section III C of reference \cite{SMA}, the order parameters 
for the modulated state are the amplitudes of the harmonics. 
The free energy that results is identical to that of Landau models in which 
the coefficient of the gradient-squared elastic terms is negative, 
necessitating the inclusion of terms quadratic in second derivatives.  
When the coefficient of the gradient-squared term vanishes, a Lifshitz
point occurs in the phase diagram \cite{Hornreich}.  
Therefore other candidates include Lifshitz-point systems such as the 
magnetic material MnP \cite{Shapira81,Becerra96}, and Langmuir monolayers 
and diblock copolymers \cite{Andelman} with modulated phases. 

Our Landau model gives a bulk temperature-field phase diagram with a closed 
loop separating the modulated phase (favored inside) from the homogeneous 
phase. 
The surprise is that the mere existence of the surface produces a surface 
state which is energetically favored within a second closed loop well 
outside the first. 
Outside the second loop, the surface state exists but it is metastable (the 
equilibrium solution is simply the homogeneous bulk state). 
The order parameter in the surface state is not a small perturbation to 
the bulk order parameter. 
The width of the state has no pronounced temperature dependence; 
in particular, it does not diverge. 
In the presence of a surface field (coupling linearly to the order 
parameter), the surface phase may still occur but the line of surface 
transitions no longer forms a closed loop.
>From the above and other evidence, our surface states are very different 
from the states considered in [1,2].

This paper is organized as follows.  
Section II presents the effective Landau-Ginzburg model and then 
describes analytical and numerical results for the bulk phase diagram. 
Section III presents analytical and numerical results for the surface 
states, in the absence of a surface field.  
Section IV shows how the surface phase diagram is modified by a surface 
field. 
Finally section V discusses the results and their possible realization.


\section{Model and bulk phase diagram}

In this section we introduce the model used in the rest of the article, 
and we study the bulk phase diagram, especially the transition line 
separating the homogeneous and modulated states.  
The bulk free energy $F_b$ is the spatial integral of the density 
${\cal F}_b$, which is the following functional of the scalar order 
parameter $\phi(x)$: 
\begin{equation}
\label{freenergy}
{\cal F}_b[\phi]=-h \phi 
+ {\textstyle{1\over2}}r \phi^2 
+ {\textstyle{1\over4}}  \phi^4
- {\textstyle{1\over2}} (\phi^{\prime})^2 
+ {\textstyle{1\over2}} (\phi^{\prime \prime})^2
\end{equation}   
where $\phi^{\prime}=d\phi/dx$. 
We have scaled the order parameter, the energy and the unit of length 
to simplify the coefficients, and so $h$ and $r$ are the rescaled ordering 
field and temperature variables respectively. 
The corresponding Euler-Lagrange equation is 
\begin{equation}
\label{euler}
\phi^{\prime\prime\prime\prime} + \phi^{\prime\prime} -h + r\phi+\phi^3=0\ .
\end{equation}
Nakanishi and Fisher\cite{Nakanishi} examined a very different model; 
the gradient-squared term appeared with a positive coefficient, the 
$(\phi^{\prime\prime})^2$ term was omitted, and surface terms were added. 
Their model, without the surface terms, applies to the usual Ising model with 
ferromagnetic interactions; 
it has only the disordered and homogeneous (ferromagnetically ordered) phases, 
and its bulk $(r,h)$ phase diagram consists of a first-order line at $h=0$ 
and $r<0$ which terminates at a critical point at $r=0$. 
The model of Equation (\ref{freenergy}), but without the $(\phi^\prime)^2$ term, 
exhibits a Lifshitz point at $h=r=0$ and a first-order line for $r<0$.

Without the bulk field $h$, the model (\ref{freenergy}) has a disordered 
phase at high temperature ($T$), a second-order transition at $r={1\over4}$ 
to a modulated phase, and a strong first-order transition at $r\approx-1.2$ 
to one of two degenerate homogeneous phases; 
the modulated phase is almost sinusoidal over its entire range, and 
its wavenumber is almost independent of $T$. 
In the $(r,h)$ plane, the modulated phase occupies a closed loop 
\cite{Coutinho80,Shapira81,Seidin}.
Outside this loop, the energetically favored phase is the homogeneous phase, 
with order parameter $\phi_0$ found from 
\begin{equation}
\label{homogeneous}
-h + r \phi_0 + {\phi_0}^3 =0 \ ;
\end{equation}   
its free-energy density is 
${\cal F}_0=-h\phi_0+{1\over2}r\phi_0^2+{1\over4}\phi_0^4$. 

Figure 1 gives the bulk phase diagram, as found for the most part by 
numerical solution of the Euler-Lagrange equation (\ref{euler}) with 
periodic boundary conditions. 
The homogeneous-modulated transition is second-order near $r={1\over4}$, but 
otherwise first-order. 
The second-order segment and the tricritical points at its ends are found 
analytically in the following. 

\begin{figure}
\psfig{figure=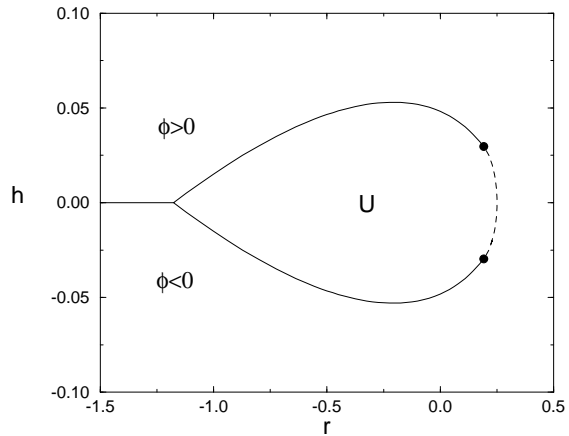,width=8cm}
\caption{ The $(r,h)$ bulk phase diagram corresponding to model (1). 
The undulating state (U) is energetically preferred inside the loop, 
and the homogeneously ordered state outside it. 
The transition between the states is either first-order (solid line)
or second-order (dashed line); 
two tricitical points (solid circles) separate the two types of transition. 
The first-order segment was found by numerical solution of the 
Euler-Lagrange equation, the second-order segment from Equation (8), 
and the tricritical points from Equation (10). 
At lower temperatures, a first-order transition at $h=0$ separates the two 
homogeneously ordered states $\phi>0$ and $\phi<0$.}
\end{figure}

We consider a spatially modulated order parameter and expand it in harmonics. 
If $q$ is the wavenumber of the modulated structure and $\epsilon$ is the 
amplitude of the leading harmonic, then the order parameter takes the form
\begin{equation}
\label{opexpansion}
\phi(x) = \phi_0 + \epsilon\cos(qx) 
+\epsilon ^2 [\phi_2\cos(2qx) +\bar \phi_2] + O(\epsilon ^3)
\end{equation}   
where $\phi_2$ and $\bar \phi_2$ are constants to be determined. 
Inserting this order parameter in the free energy (\ref{freenergy}) and 
integrating over a period, one finds the following expansion of the free 
energy (per unit volume): 
\begin{equation}
\label{feexpansion}
\langle{\cal F}_b\rangle={\cal F}_0 
+ \epsilon^2{\cal F}_2 + \epsilon^4{\cal F}_4 + O(\epsilon^6)
\end{equation}   
with coefficients 
\begin{equation}
\label{F2}
{\cal F}_2 = {\textstyle{1\over4}}  (r+3\phi_0 ^2 -{\textstyle{1\over4}}) \ ,
\end{equation}   
\begin{eqnarray}
\label{F4}
{\cal F}_4 & = & 
  {\textstyle{1\over4}} (r+3\phi_0 ^2 +2)\phi_2 ^2 
+ {\textstyle{1\over2}} (r+3 \phi_0 ^2) {\bar \phi_2 ^2} \nonumber \\
& + & 
  {\textstyle{3\over2}} \phi_0 {\bar \phi_2} 
+ {\textstyle{3\over4}} \phi_0 \phi_2 
+ {\textstyle{3\over32}} \ .
\end{eqnarray}   
The free energy has already been minimized with respect to the wavenumber 
$q$, giving $q=\sqrt{1/2}+O(\epsilon^2)$. 
The homogeneous phase is unstable to a modulated perturbation when 
${\cal F}_2<0$. 
Provided that ${\cal F}_4>0$ then, a second-order transition occurs at 
\begin{equation}
\label{critical-line}
h= \pm {2 \over{3{\sqrt 3}}} 
{\left({1 \over {8}} +r\right) \sqrt{{1 \over {4}} -r}} \ .
\end{equation}   
When ${\cal F}_4$ is negative, the transition to the modulated phase is 
first-order. 
To find the tricritical points separating the continuous and first-order 
segments, we minimize ${\cal F}_4$ with respect to $\phi_2$ and $\bar \phi_2$ 
and then set the result equal to zero. 
On the line ${\cal F}_2=0$, ${\cal F}_4$ is minimized by 
$\phi_2=-2\phi_0/3 $ and $\bar\phi_2=-6\phi_0$; 
the minimum value is 
\begin{equation}
\label{F4=0}
{\cal F}_4={\textstyle{3\over32}} -{\textstyle{19\over4}} \phi_0 ^2 \ ,
\end{equation}   
and so the two tricritical points are located at 
\begin{equation}
\label{tcp}
r=29/152 ~~~,~~~ h=\pm \sqrt{6/19^3} \ .
\end{equation}   
%
%


\section{Surface phase diagram}

In this section we consider the surface phase diagram of the model 
(\ref{freenergy}) for a semi-infinite system, with no surface field. 
The presence of the surface generally produces states localized near the 
surface, and the states are energetically favored in part of the phase 
diagram. 
We studied the surface states in the region where the bulk phase is 
homogeneous, and examined their transitions with varying temperature and 
the external field. 
Only a cursory examination was made in the region where the bulk is modulated; 
in this region, we found many solutions of the Euler-Lagrange equation, 
so many that a detailed analysis was felt unjustified at this time. 
That is, surface states and surface phase transitions may exist inside the 
bulk modulated loop, but have not been studied. 

We consider a system occupying the half-space $x\geq0$, and we assume that 
the order parameter depends only on $x$. 
The bulk energy $F_b$ is found by integrating the density of (\ref{freenergy}). 
In this section, we treat the surface very simply, by assuming that it merely 
terminates the bulk; 
we thus take free boundary conditions at the surface. 
In section IV, however, we assume that the surface also applies a local 
ordering field $h_s$; 
then the total energy is $F_b+F_s$, where 
\begin{equation}
\label{surface.fe}
F_s = -h_s \phi_s 
\end{equation}   
and $\phi_s$ is the order parameter at $x=0$. 
The general boundary conditions are then 
\begin{equation}
\label{bcbcbc}
\phi_s^\prime + \phi_s^{\prime\prime\prime}-h_s=0\ ,
\ \ \ \ \phi_s^{\prime\prime}=0 \ . 
\end{equation}

We solved the Euler-Lagrange equation (\ref{euler}) numerically
subject to the boundary conditions (\ref{bcbcbc}). 
This equation can have many solutions, depending on the bulk field $h$ and 
the temperature variable $r$. 
Figure 2 gives the surface phase diagram for $h_s=0$, as found from 
examining these solutions. 
The surface states are energetically favorable inside the outer loop of the 
figure (with the qualification noted above), and the homogeneous states 
outside; 
the surface states exist (as solutions of the Euler-Lagrange equation) 
outside this loop but are only metastable there. 
An interesting feature is that the surface orders at $r=1$ for $h=0$, 
but the bulk orders only at $r={1\over4}$. 
Many more surface states were found at lower temperatures, but they were 
always metastable. 
\begin{figure}
\psfig{figure=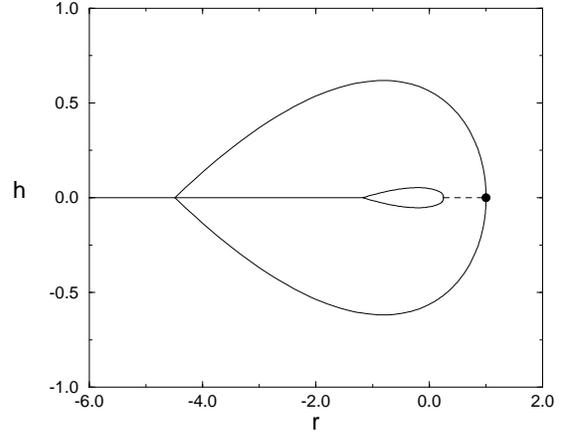,width=8cm}
\caption{The $(r,h)$ surface phase diagram for $h_s=0$. 
The surface states are energetically favorable inside the outer loop (with 
the qualification noted in the text), and the homogeneous states outside.  
The transition is first-order except at the isolated point $r=1$, $h=0$ 
(indicated by a dot) where it is continuous.  
The inner loop is the loop of Figure 1 (the scale precludes display of 
details). 
The leftmost of the three horizontal lines at $h=0$ is the bulk transition 
between the ordered states $\phi>0$ and $\phi<0$. 
At the rightmost line (dashed), the surface state changes discontinuously 
and the bulk state continuously.
At the middle line, both surface and bulk states change discontinuously.} 
\end{figure}

Figure 3 shows a typical profile of the surface state in the ordered region, 
for a small and negative bulk field (to break the symmetry) and $h_s=0$. 
The order parameter decays to the bulk value (which is negative) far from 
the surface, but it is large and positive near it; 
the overshooting and the damped oscillations result from a complex decay 
constant, as shown below. 
Correspondingly, when the bulk field is positive, the order parameter 
of the surface state is negative near the wall and then decays to the 
positive bulk value. 
Thus at $h=0$ there is a first-order transition at which the surface state 
changes sign. 

\begin{figure}
\psfig{figure=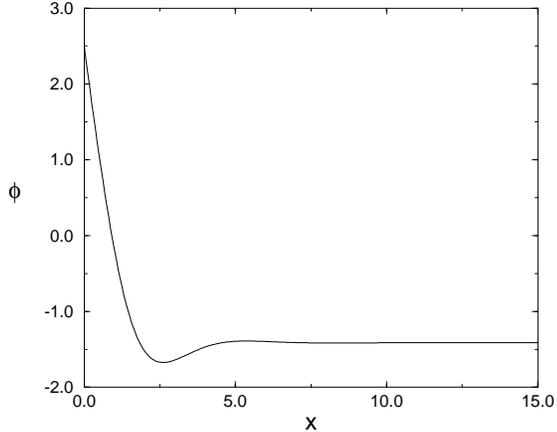,width=8cm}
\caption{Order parameter $\phi(x)$ of the surface state for parameters 
$r=-2$, $h=0^-$ and $h_s=0$. 
The order parameter is large and positive near the surface; 
it crosses zero and then decays to the bulk value for large $x$.} 
\end{figure}

To provide an analytical understanding of these numerical results and also 
those of the next section, we present the following stability analysis of 
the homogeneous bulk state. 
The analysis is valid when the deviation of the order parameter from 
the bulk value is small. 

The order parameter is written as $\phi = \phi_0 + \psi$, where $\phi_0$ is 
given by (\ref{homogeneous}) and $\psi$ is the deviation. 
The free-energy density ${\cal F}={\cal F}_b - {\cal F}_0$ associated with 
$\psi$ is 
%
%
\begin{equation}
\label{freenergy.psi}
{\cal F}=
 {\textstyle{1\over2}} (r+3 {\phi_0}^2) \psi^2 
+\phi_0 \psi^3
+{\textstyle{1\over4}} \psi^4 
-{\textstyle{1\over2}} (\psi^{\prime})^2 
+{\textstyle{1\over2}} (\psi^{\prime \prime})^2 \ .
\end{equation}   
The energy is minimized by an order parameter $\psi$
which satisfies the Euler-Lagrange equation
\begin{equation}
\label{euler.equation}
(r+3 {\phi_0}^2) \psi +3 \phi_0 {\psi^2} + \psi^3
+\psi^{\prime \prime} + \psi^{\prime \prime \prime \prime} =0 \ .
\end{equation}   
To prepare for the next section, we include also the surface free energy 
(\ref{surface.fe}).
The boundary conditions are then 
\begin{equation}
\label{boundary.equations}
\psi^{\prime}(0) +\psi^{\prime \prime \prime}(0) -h_s=0\ , 
\ \ \ \ \psi^{\prime \prime}(0)=0 \ .
\end{equation}
For $h_s=0$, the homogeneous bulk state $\psi=0$ is clearly a solution of 
Equations (\ref{euler.equation}) and (\ref{boundary.equations}). 
This solution is stable over some region of the $(r,h)$ plane, but it becomes 
unstable at the transition to the bulk modulated state. 

To study the surface states, we solve Equations (\ref{euler.equation}) and 
(\ref{surface.fe}) perturbatively in $\psi$. 
The expansion starts from the solution 
\begin{equation}
\label{linear.solution}
\psi_1(x) = A e^{-\alpha x} + A^{*} e^{- \alpha^* x}
\end{equation}   
of the linearized Equation (\ref{euler.equation}). 
The amplitude $A$ and the decay constant $\alpha$ are both complex; 
the latter (with positive real part) is found from 
\begin{equation}
\label{alpha}
\alpha^2={\textstyle{1\over2}}(-1+i\gamma)
\end{equation}   
where $i= \sqrt{-1}$ and $\gamma = [4(r+3 \phi_0^2)-1]^{1/2}$. 
The condition $\psi_1^{\prime\prime}(0)=0$ in (\ref{boundary.equations}) 
gives the amplitude $A$ in terms of $m=\psi(0)$ as 
\begin{equation}
\label{A}
A={m \over 2}\left(1 - {i \over \gamma}\right) \ .
\end{equation}   
It is convenient to take $m$ as the expansion parameter. 

The solution (\ref{linear.solution}) gives the free energy to order $m^2$. 
In order to obtain the free energy to the required order ($m^4$), one must 
find the higher-order contributions to $\psi$. 
Let $\psi = \psi_1 + \psi_2$, where $\psi_2$ is the nonlinear part of $\psi$. 
Inserting this form in Equation (\ref{euler.equation}), using 
(\ref{linear.solution}) and keeping terms to $m^3$, one finds
\begin{eqnarray}
\label{psi2}
\psi_2(x) & = & B_1 e^{-2 \alpha x} 
+ {\textstyle{1\over2}} B_2 e^{-(\alpha + \alpha^*)x} +  C_1 e^{-3 \alpha x}
\nonumber \\
& + & C_2 e^ {-(2\alpha + \alpha^*)x} + D e^{- \alpha x} + c.c.
\end{eqnarray}  
with $\psi_2(0)=\psi_2^{\prime\prime}(0)=0$ and coefficients 
\begin{eqnarray}
B_1 & = & -\left(3+5\sqrt3 i\right) \phi_0 m^2 / 126 \ ,\nonumber \\ 
B_2 & = & -2 \phi_0 m^2 /3                           \ ,\nonumber \\ 
C_1 & = & -\left(27 + 11 \sqrt3 i\right) m^3 / 13104 \ ,\nonumber \\ 
C_2 & = &  \left( 3 +  2 \sqrt3 i\right) m^3 /    84 \ ,\nonumber \\ 
  D & = & \left( {5\over 14} - {19\over 126} \sqrt3 i\right) \phi_0 m^2 
         -\left( {7\over208} + {47\over4368} \sqrt3 i\right) m^3 
                                                     \ .\nonumber 
\end{eqnarray}   
On using the result $\psi=\psi_1+\psi_2$ in the free energy 
(\ref{freenergy.psi}) and integrating over $x$, one finds that the free 
energy of the surface state (per unit area) is given by
\begin{equation}
\label{F}
F=-h_s m +a_2 m^2 
+{\textstyle{2\over9}} \phi_0 m^3 
+{\textstyle{3\over56}} m^4 + O(m^5) 
\end{equation}   
where
\begin{equation}
a_2=-\left(r+3 \phi_0^2\right){i \over 2\gamma}{{\alpha^3 - \alpha^{*3}}
\over \alpha \alpha^*} \ .
\end{equation}   
The amplitude of the surface structure is determined by minimizing
the free energy with respect to $m$ for given surface field $h_s$.
This amounts to satisfying the first condition in Equation 
(\ref{boundary.equations}). 
We now use the free energy (\ref{F}) to discuss the surface phase 
diagram in the region where $m$ is small. 

Consider first the case $h_s=0$. 
For zero bulk field $h$, $\phi_0=0$ and there is no surface state when 
$a_2>0$ (that is, $\psi=0$).
Setting $a_2=0$, one finds a continuous transition at $r=1$ from the 
disordered bulk state $m=0$ to a surface state with $m\neq0$; 
this is the second-order point at the right of Figure 2. 
For field $h\neq0$, the bulk order parameter $\phi_0$ is also non-zero 
and the free-energy expansion (\ref{F}) has a term in $m^3$; 
this cubic term gives a first-order transition to the surface state, again 
as found numerically. 
Near the point $(r=1, h=0)$, $m$ is small and the transition line can be found 
approximately from the free-energy expansion ({\ref{F}). 
Away from this point, however, the full free energy must be minimized 
numerically; 
Figure 2 gives the resulting $(r,h)$ surface phase diagram for $h_s=0$. 

\section{Effect of a surface field} 

We consider now the surface phase diagram for non-zero surface field $h_s$. 
Positive $h_s$, for example, tends to increase the order parameters of 
{\it all} states in the region near the surface. 
The new feature is that the Euler-Lagrange equation must now be solved 
numerically for what were homogeneous bulk states at $h_s=0$;  
for lack of a better term, we refer to these surface-field-modified bulk 
states simply as bulk states. 
Figures 4 and 5 give parts of typical phase diagrams for $h_s>0$, as found by 
numerical solution of Equation (\ref{euler}), subject to the boundary 
conditions (\ref{bcbcbc}). 

Figure 4 shows the high-temperature part of the phase diagram for 
$h_s=10^{-4}$. 
The surface field breaks the transitions of Figure 2 into two first-order 
lines at which the surface state changes discontinuously. 
In the region bounded by the upper line and the left vertical (where the bulk 
field $h$ is positive), the order parameter of the surface state is negative 
at the boundary $x=0$ ($\phi_s<0$). 
The lower line ends at a second-order point.
Below this point there is a first-order transition between the 
paramagnetic state and the surface state with $\phi_s>0$, 
while above it the two states are indistinguishable. 
The free-energy expansion (\ref{F}) can be used to find this point to leading 
order in $h_s$; 
the result is 
\begin{eqnarray}
\label{critical.point}
a_2 & = & {3 \over 2} {\left({3 \over 14}\right)^{1/3}} h_s^{2/3} 
                                              \ , \ \nonumber \\
\phi_0 & = & - {9 \over 2} {\left({3 \over 14}\right)^{2/3}} h_s^{1/3} \ ,
\end{eqnarray}   
in good agreement with the numerical results. 

\begin{figure}
\psfig{figure=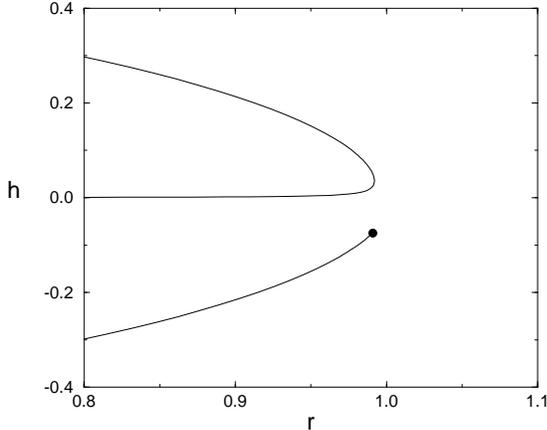,width=8cm}
\caption{The high-temperature part of the $(r,h)$ surface phase 
diagram for $h_s=10^{-4}$.  
Both transition lines are first-order. 
The lower line ends at a second-order point marked by the solid circle 
(see text).}
\end{figure}

Figure 5 shows the low-temperature part of the phase diagram for 
$h_s=10^{-1}$. 
Paradoxically, a positive surface field cooperates, rather than competes, 
with a negative bulk field to enhance the stability of the lower surface 
state (and it competes with a positive bulk field for the other). 
These effects occur because the order parameter of the surface state changes 
sign (as seen in Figure 3). 

\begin{figure}
\psfig{figure=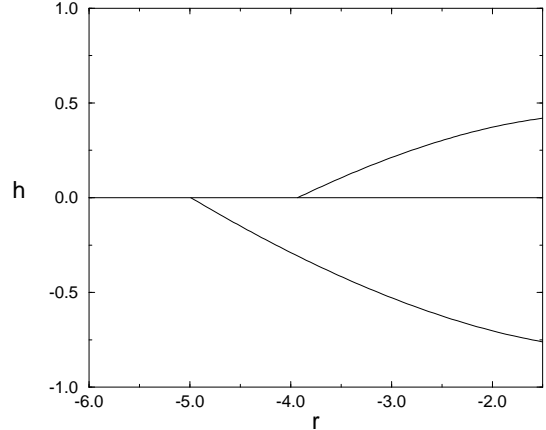,width=8cm}
\caption{The low-temperature part of the $(r,h)$ surface phase 
diagram for $h_s=10^{-1}$.  
All transitions are first-order. 
The surface field enhances the stability of the lower surface state, for 
which $\phi(0)>0$ and $\phi(\infty)<0$, and decreases the stability of the 
other. 
The leftmost segment of the horizontal line at $h=0$ represents the 
transition between the two bulk states; 
these states are not homogeneous in the presence of the surface field. 
The other segments describe bulk-driven instabilities of the surface states; 
for example, the lower surface state cannot exist for $h>0$.} 
\end{figure}
%
%
 

\section{Discussion}

We have developed and analyzed a model to describe the effect of a substrate 
(or a free surface) on a material which has a bulk phase transition between 
homogeneous and modulated states.
Modulated states tend to form because the free energy of the model contains 
a term, quadratic in first derivatives of the order parameter, which has a 
negative coefficient.  
We treated the surface first as simply terminating the bulk, and then in 
addition as supplying a surface field coupling linearly to the order 
parameter. 

The important new result of our analysis is the quite unexpected existence 
of solutions localized at the surface, solutions which exist even if the 
surface field is zero. 
These solutions are energetically favored for temperature and field values 
that are outside but not too far from the closed loop within which the 
modulated bulk state is stable.  
When the surface field differs from zero, the loop breaks apart (as shown 
in Figures 4 and 5). 

We now turn our attention to the applicability of our results to cholesteric 
liquid crystals in a field \cite{SMA}. 
It is obviously desirable to estimate the conditions of chirality, 
temperature, field, and surface interactions for which the surface states 
should be observable.  
To do this, we should examine the relationship between the variables of the 
theory and the experimental variables, by comparing the expressions for 
${\cal F}_2$ and ${\cal F}_4$ in section II of this paper with the analogous 
expressions in section III-C of \cite{SMA}.  
It is reasonable however, and far simpler, to expect the loop regions to 
scale by the same factors; 
this should be true independent of the strength of surface interactions.  
From Figures 1 and 2, the outer (surface-state) loop extends over the range
$-4.5 \alt r < 1$ while the inner loop extends over 
$-1.2 \alt r < {1\over4}$, about a factor of four. 
Accordingly, we estimate the surface-state region to be four times the 
size of the undulating-state region in temperature. 
From \cite{SMA}, the undulating state should occur for intrinsic pitches in 
the range of $1260-630$ nm, at electric fields of the order of a few 
hundreds of kV/cm, or magnetic fields of roughly $40$ T;  
the temperature width was estimated to be a few tenths of a degree.  
These conditions are very difficult to achieve and account for the fact that 
the undulating state has not yet been observed, although some groups plan 
to attempt the experiments.  
The surface-state region is expected to be one degree wide. 
Techniques sensitive to birefringence near the surface, such as 
Brewster-angle ellipsometry \cite{Lucht1,Lucht2}, may be able to detect the 
surface states.

The surface states should appear in incommensurate systems where the 
modulated phase is driven by a negative gradient-squared term. 
Conditions may be favorable in magnetic Lifshitz-point materials like MnP, 
or in Langmuir monolayers or diblock copolymers. 
Other systems in which a modulated phase is driven by a negative 
gradient-squared term are sodium nitrite and thiourea \cite{cummins}}; 
related systems are quartz and berlinite, but for these the modulated phase 
is two-dimensional.
%
%
%
%
\acknowledgments  
We thank R. Seidin for helpful discussion of the bulk phase diagram and
R. C. Desai for helpful comments.
This research was supported by the National Science Foundation under Science 
and Technology Center ALCOM Grant No. DMR 89-20147, 
the Einstein Center for Theoretical Physics, 
the Inter-University High Performance Computation Center (Tel Aviv), 
the Natural Sciences and Engineering Research Council of Canada, and the 
Meyerhoff Foundation.

%
%
%
%
  
\end{document}